\documentclass[aip,apl,reprint,superscriptaddress]{revtex4-1}
\usepackage{graphicx}
\usepackage{textcomp}
\usepackage{hyperref}

\hypersetup{
    colorlinks=true,
    linkcolor=black,
    citecolor=black,
    filecolor=black,
    urlcolor=black,
}

\begin{document}


\title{Bandgap widening by disorder in rainbow metamaterials}


\author{Paolo Celli}
\affiliation{Department of Civil, Environmental, and Geo- Engineering, University of Minnesota,\\ Minneapolis, MN 55455, USA}
\affiliation{Division of Engineering and Applied Science, California Institute of Technology,\\ Pasadena, CA 91125, USA}
\author{Behrooz Yousefzadeh}
\affiliation{Division of Engineering and Applied Science, California Institute of Technology,\\ Pasadena, CA 91125, USA}
\author{Chiara Daraio}
\affiliation{Division of Engineering and Applied Science, California Institute of Technology,\\ Pasadena, CA 91125, USA}
\author{Stefano Gonella}
\email{sgonella@umn.edu}
\affiliation{Department of Civil, Environmental, and Geo- Engineering, University of Minnesota,\\ Minneapolis, MN 55455, USA}


\begin{abstract}
\vspace{5px}
\normalsize{\textbf{This article may be downloaded for personal use only. Any other use requires prior permission of the author and AIP Publishing. This article appeared in}: \emph{Appl. Phys. Lett.} {\bf 114}, 091903 (2019) \textbf{and may be found at}: \url{https://doi.org/10.1063/1.5081916}}
\vspace{15px}

Stubbed plates, i.e., thin elastic sheets endowed with pillar-like resonators, display subwavelength, locally-resonant bandgaps that are primarily controlled by the intrinsic resonance properties of the pillars. 
In this work, we experimentally study the bandgap response of a tunable heterogeneous plate endowed with reconfigurable families of pillars. We demonstrate that, under certain circumstances, both the spectrum of resonant frequencies of the pillars and their spatial arrangement influence the filtering characteristics of the system. Specifically, both spatially graded and disordered arrangements result in bandgap widening. Moreover, the spectral range over which attenuation is achieved with random arrangements is on average wider than the one observed with graded configurations.

\end{abstract}

\keywords{Metamaterials, Rainbow trap, Tunability, Modularity, Disorder, Bandgap widening}

\maketitle


Due to their peculiar mesoscale architectures, metamaterials are capable of manipulating waves in the \emph{subwavelength} regime~\cite{Liu_SCIENCE_2000, Zhu_NAT-PHYS_2011, Zhang_PRL_2011, Lemoult_NAT-PHYS_2013, Hladky_APL_2013, Zhu_PRL_2016}---when the wavelengths are much larger than the characteristic microstructural length scales of the medium. These phenomena are typically achieved via homogeneous arrays of resonators, but some effects require the coexistence of \emph{heterogeneous} populations of resonators with different spectral characteristics. For example, subwavelength waveguiding has been achieved by frequency upshifting selected resonators along a desired waveguide path~\cite{Lemoult_NAT-PHYS_2013, Addouche_APL-ADV_2014, Gao_APL_2016, Kaina_arXiv_2016}, and topological effects have been observed in systems comprising hexagonal arrangements of different resonator types~\cite{Pal_NJoP_2017, Chaunsali_arXiv_2018}. Similarly, rainbow trapping requires arrays of resonators with different characteristics, where each type of resonator distills a selected frequency from a broadband input signal~\cite{Tsakmakidis_NATURE_2007, Zhu_SCIREP_2013, Kroedel_EML_2015, Zhao_SCIREP_2015, Cardella_SMS_2016, Tian_SCIREP_2017, Colombi_SCIREP_2017, Guo_NJoP_2017}. Heterogeneity introduces an additional degree of freedom for metamaterial design that stems from the \textit{spatial} arrangement of the resonating units. Determining whether this spatial arrangement has any influence on the wave control capabilities of the system requires understanding how neighboring resonators are coupled by the wave-carrying substrate medium. If such influence is indeed observed, it is of practical interest to determine which spatial configuration maximizes the desired effect.

\begin{figure} [!htb]
\centering
\includegraphics[scale=1.4]{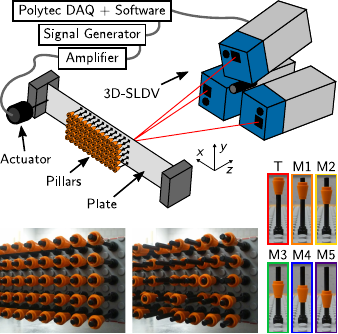}
\caption{Experimental setup (top). The bottom images illustrate two configurations achievable through different selections and arrangements of the six pillar types shown on the right (T to M5), obtained by sliding the conical brick down the rod by discrete increments.} 
\label{fig:setup}
\end{figure}
\begin{figure*} [!htb]
\centering
\includegraphics[scale=1.4]{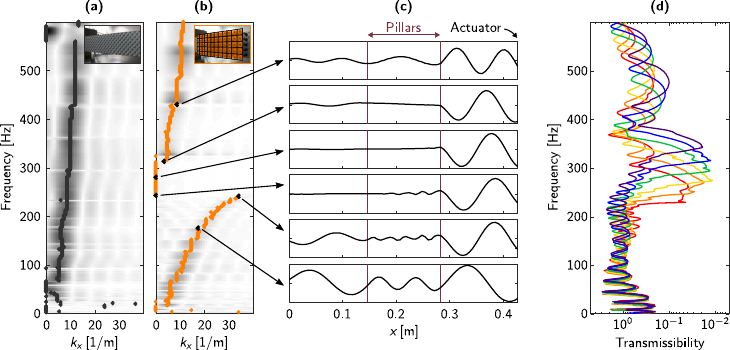}
\caption{(a,b) Experimentally-reconstructed dispersion relations { ($k_x$ is the wave number along the $x$-direction)} of the bare baseplate and of the baseplate with 5$\times$12 identical M1 pillars, respectively. (c) Deformed shapes for the (b) case, measured at the centerline of the strip for frequencies below, inside, and above the gap. (d) Transmissibilities of specimens of identical pillars, matching the six types color-coded in Fig.~\ref{fig:setup}.}
\label{fig:uni}
\end{figure*}
In this work, we study the effects of the spatial arrangement of local resonators on the bandgap characteristics of an elastic metamaterial, with special emphasis on bandgap widening. Our medium of choice is a stubbed plate in which the plate substrate behaves as the wave-carrying medium, and arrays of surface pillars act as resonators~\cite{Wu_2008, Pennec_2008, Oudich_PRB_2011, Assouar_APL_2012, Casadei_JAP_2012, Achaoui_JAP_2013, Rupin_PRL_2014, Pourabolghasem_JAP_2014, Celli_APL_2015} (Fig.~\ref{fig:setup}). It is documented that graded arrangements of heterogeneously-tuned resonators display wider bandgaps than their homogeneous counterparts~\cite{Kroedel_EML_2015}. Here, our aim is to highlight the role played by the compliance of the wave-carrying medium in controlling the bandgap response. When the plate is significantly more compliant than the pillars, the wave transmission characteristics become more sensitive to the spatial arrangement of the resonators. Thus, even heterogeneous arrangements that differ from each other solely in terms of their spatial characteristics result in appreciably different bandgap responses. Specifically, we intend to show that random  arrangements systematically widen the bandgaps with respect to nominally similar graded configurations.

In order to achieve the versatility required to test multiple architectures within a single reconfigurable specimen, we design a modular~\cite{Wu_JIMSS_2015, Lee_SCIREP_2016, Memoli_NATCOMM_2017} stubbed plate with tunable resonators. We arrange arrays of LEGO\textsuperscript{\textregistered} bricks on a thin baseplate strip (Fig.~\ref{fig:setup}), following a testing paradigm that we previously introduced for homogeneous locally-resonant phononic crystals~\cite{Celli_APL_2015}. The specimen features 12$\times$5 pillars arranged according to a square lattice in the central section of the strip. Each resonator can assume one of the six discrete inertial configurations shown at the bottom-right of Fig.~\ref{fig:setup}, obtained by sliding the conical tip down the rod by discrete increments of $\delta h = 3.25\,\mathrm{mm}$. This manual tuning alters the effective inertial characteristics of a pillar and thereby its natural frequencies, and is conceptually similar to strategies discussed by other authors for Bragg bandgap tunability~\cite{Goffaux_PRB_2001, Romero-Garcia_JPD_2013, Thota_PRB_2017}. Two examples of different pillar arrangements are shown at the bottom of Fig.~\ref{fig:setup}. The plates are excited with a pseudorandom waveform prescribed by a shaker to establish standing flexural wave patterns over a broad spectrum of frequencies, and their out-of-plane response is recorded with a 3D Scanning Laser Doppler Vibrometer (3D-SLDV). More details on the setup are discussed in the \emph{Supplementary Material} (SM) section.

The effect of introducing arrays of pillars is to open locally-resonant bandgaps in the phonon band structure. This is clear when comparing the dispersion relation (reconstructed from experimental data) of the bare baseplate, which features a single flexural mode in the range of interest, against that of the baseplate with all pillars of the M1 type, where the mode is split and a hybridization gap arises between 242 and 311 Hz (Fig.~\ref{fig:uni}a,b). Here the dotted lines that follow the dispersion branches are obtained by tracking the maxima of the spectral amplitude (the underlying grayscale colormap). More details on the band diagram reconstruction are given in the SM. The flexural deflection shapes recorded along the strip's centerline (Fig.~\ref{fig:uni}c) reveal some peculiar aspects of the physical mechanisms responsible for this gap. These observations are crucial to properly interpret the results discussed in the rest of this work. {It can be noticed that, at the onset of the gap, the region of the plate below the pillars undergoes flexural deformation, whose amplitude is not negligible with respect to the maximum plate deflection and with respect to the motion of the pillars. This suggests that, in our system, what effectively resonates are not the pillars taken as stand-alone beam-like structural elements forced at their base (as in other works featuring similar platforms~\cite{Rupin_PRL_2014, Williams_PRB_2015}), but rather the whole unit cell comprising plate and pillar. In particular, each pillar's main contribution is to add rotatory inertia to one resonating unit.} We conjecture that {this resonant behavior is} due to the specific landscape of mechanical properties in our specimen, where the plate has low relative stiffness with respect to the pillars, which establishes strong coupling between neighboring pillars through the plate substrate. For completeness, we analyzed the response of different uniform configurations, each featuring one of all the possible resonator types. The measured transmissibilities indicate a trend of shifted and partially overlapping bandgaps (Fig.~\ref{fig:uni}d). As the tip mass is slid down the pillars and their effective inertia is decreased, the onsets of the bandgaps shift towards higher frequencies (see dispersion curves in the SM section).

\begin{figure} [!htb]
\centering
\includegraphics[scale=1.4]{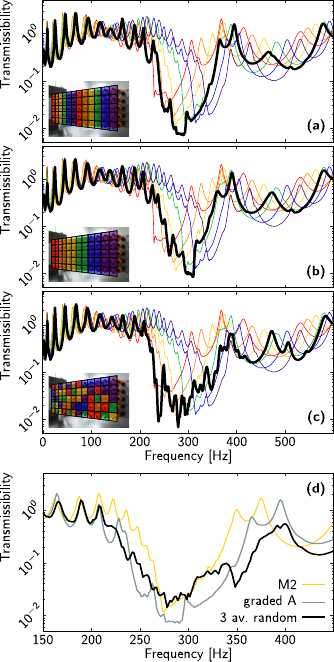}
\caption{Influence of the spatial arrangement of resonators on wave attenuation. The resonators are divided into six groups of ten units each, implementing the six configurations shown in Fig.~\ref{fig:setup}. { (a-b) Transmissibilities for graded configurations (the pattern in (a) is called graded A and the one in (b) graded B) and (c) spatially randomized configurations, marked by thick black lines; thin color-coded lines refer to monochromatic configurations.} (d) Comparison of bandgaps for uniform, graded and random (3 averages) configurations.}
\label{fig:bgap}
\end{figure}

Locally resonant bandgaps are usually narrow and therefore impractical to design mechanical filters that are effective against broadband excitations. This limitation has inspired numerous widening strategies, among which we recall trampoline effects~\cite{Bilal_APL_2013}, rainbow trapping~\cite{Kroedel_EML_2015, Banerjee_JAP_2017}, bandgap adjoining~\cite{Coffy_JAP_2015, Oh_APL_2016}, and disorder-based methods involving Anderson localization phenomena~\cite{Thorp_SMS_2001, Sainidou_PRL_2005, Richoux_2009}. Here, we investigate the behavior of heterogeneous populations of resonators with emphasis on the dependence of the bandgap width upon the spatial arrangement of the resonators. To this end, we compare the performance of three classes of configurations: i) uniform arrangements, ii) graded arrangements of heterogeneous resonators, and iii) random populations of heterogeneous resonators. The measured transmissibilities of a few representative configurations are shown in Fig.~\ref{fig:bgap}. 
In each subfigure, the thick black line represents the transmissibility for the arrangement shown in the corresponding inset; all the configurations feature 10 resonators of each type---T, M1, M2, M3, M4, M5 denoted by the color coding introduced in Fig.~\ref{fig:setup}. {The results for two graded architectures with different gradient patterns (the pattern in Fig.~\ref{fig:bgap}a, called graded A, and the one in Fig.~\ref{fig:bgap}b, labeled graded B) clearly highlight a widening of the bandgap with respect to their monochromatic counterparts.} In both cases, the total bandgap spans the frequency interval encompassing the individual gaps of three monochromatic configurations. This result is a manifestation of the rainbow trapping effect~\cite{Kroedel_EML_2015, Zhao_SCIREP_2015, Cardella_SMS_2016}. {Interestingly, we observe that the graded A configuration produces a wider and deeper gap than the graded B one, suggesting that the performance of graded architectures is influenced by the period of the spatial arrangement.} 
The transmissibility of a representative spatially-disordered arrangement (MATLAB-generated) is shown in Fig.~\ref{fig:bgap}c. The most distinctive morphological difference brought about by randomization is that the bandgap is wider than its graded and homogeneous counterparts---stretching here over the frequency interval spanned by five individual bandgaps. The reliability of this observation is confirmed by averaging three random realizations (Fig.~\ref{fig:bgap}d, black line). These experimental results (later corroborated by numerical results) lead to the conclusion that randomization causes attenuation over a wider frequency range, but this comes at the expense of the attenuation amplitude, which decreases with respect to the graded case. A similar effect is well documented in non-resonant disordered systems\cite{Lin_AMR_1996}.

\begin{figure} [!htb]
\centering
\includegraphics[scale=1.4]{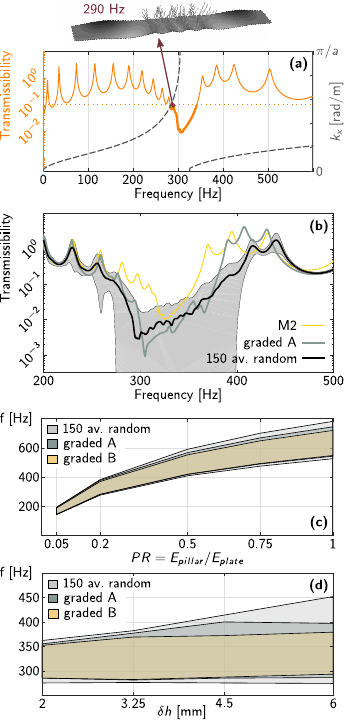}
\caption{Numerical simulations. (a) Dispersion relation (dashed gray line) and transmissibility (orange line) for the M1 homogeneous configuration. We consider everything below the dotted orange line (transmissibility $<10^{-1}$) as bandgap. The inset depicts a deflection shape right before the bandgap onset. (b) Comparison of bandgaps for uniform, graded and random (average of 150 realizations) configurations. The gray area is the standard deviation of the black curve. (c,d) Bandgap width as a function of the plate ratio ($PR$), and of the increment by which the mass is slid down the pillars ($\delta h$), respectively. Default values are $\delta h = 3.25$ mm, $PR=0.2$.}
\label{fig:num}
\end{figure}

To substantiate our experimental findings, and to explore the robustness of these results against variations in the characteristics of plate and pillars, we perform finite element simulations, carried out in Abaqus/Standard. The plate is discretized using 3D shell elements. The pillars are modeled as Timoshenko beams that are assumed to be perfectly anchored to the plate at a single point. The conical sliding tip is assumed to be a point mass. 
The numerical model captures all the important qualitative features of the experimental results, albeit without matching quantitatively the frequencies of the experimental bandgaps.
For example, the dispersion relation and the transmissibility curve (dashed gray lines and orange line in Fig.~\ref{fig:num}a, respectively) for the M1 homogeneous configuration feature trends that closely resemble the experimental results in Fig.~\ref{fig:uni}b,d. The deflection shape at the onset of the bandgap (inset in Fig.~\ref{fig:num}a) highlights that the resonance is characterized by an undulatory motion of the plate, also consistent with the experiments (Fig.~\ref{fig:uni}c). An advantage of numerical simulations is that we can compare a significantly larger set of realizations of the disordered metamaterial to infer the statistical behavior of the ensemble. Fig.~\ref{fig:num}b shows the average transmissibility curve obtained by averaging 150 disordered realizations. These results corroborate the previous conclusion (Fig.~\ref{fig:bgap}d) that the locally-resonant bandgap of the random configuration is on average wider than the bandgap of the graded configurations. More detailed comparisons are provided as SM.


We leverage the numerical platform to further explore the influence of some key system parameters on the bandgap behavior. In particular, we explore the effect of i) the relative stiffness between plate and pillars, ii) the spacing between the resonance frequencies of the pillars. For this parametric study, we focus on the bandgap width, specifically on how it compares between the random and the two graded configurations.
Fig.~\ref{fig:num}c shows that stiffening the plate (as parameterized by $P\!R$, the ratio between the elastic modulus of the pillar and that of the plate) will widen the bandgap for all three configurations, as well as shift them towards higher frequencies. The latter aspect provides further evidence that the plate is part of the resonating unit.
Fig.~\ref{fig:num}d shows the bandgap width as a function of $\delta h$, which is the increment by which the conical mass is slid down the pillars ($\delta h$ was set to the default 3.25 value in the previous discussion). Increasing $\delta h$ results in a wider separation between the bandgaps of the six configurations made of uniform pillar arrangements, and therefore widens the total bandgaps for both the graded and the random configurations \cite{blah}. However, it is clear that the widening effect is more pronounced for random configurations. It is nonetheless noted that, although random configurations have the widest bandgap on average, the widening does not necessarily occur for all realizations, as evidenced by the standard deviation (gray shading in Fig.~\ref{fig:num}b). 
From this parametric analysis we conclude that the disorder-induced widening of bandgap is robust to small changes in structural and material properties of the metamaterial, and occurs over a wider frequency range than the bandgap produced by grading the resonators.

In conclusion, this work takes advantage of highly-reconfigurable LEGO\textsuperscript{\textregistered}-brick-based stubbed plates to highlight the potential advantages of mechanical metamaterials featuring heterogeneous populations of resonators as broadband filters and showcases the importance of the spatial arrangement of the resonators. In particular, we have realized and tested a tunable elastodynamic rainbow trap and we have shown that the interplay between heterogeneity in the resonators' characteristics and spatial disorder results in a widening of the filtering effects compared to the more conventional spatially-ordered rainbow material concepts. 


\section*{SUPPLEMENTARY MATERIAL}
See supplementary material for a more detailed account on the experimental setup, the numerical model, and for additional results.

\begin{acknowledgments}
S.G.\ and P.C.\ acknowledge support from the National Science Foundation (CMMI-1266089). P.C.\ acknowledges support from the U.\ of Minnesota Doctoral Dissertation Fellowship. B.Y.\ acknowledges support from the Natural Science and Engineering Research Council of Canada through a postdoctoral fellowship. C.D.\ and B.Y.\ acknowledge partial support from the National Science Foundation (EFRI-1741565). We thank W. Zhang, N. Bausman, M. Turos, A. Palermo, V. Tournat and A. Cebrecos for their help and fruitful discussions. 
\end{acknowledgments}


\begin{thebibliography}{44}%
\makeatletter
\providecommand \@ifxundefined [1]{%
 \@ifx{#1\undefined}
}%
\providecommand \@ifnum [1]{%
 \ifnum #1\expandafter \@firstoftwo
 \else \expandafter \@secondoftwo
 \fi
}%
\providecommand \@ifx [1]{%
 \ifx #1\expandafter \@firstoftwo
 \else \expandafter \@secondoftwo
 \fi
}%
\providecommand \natexlab [1]{#1}%
\providecommand \enquote  [1]{``#1''}%
\providecommand \bibnamefont  [1]{#1}%
\providecommand \bibfnamefont [1]{#1}%
\providecommand \citenamefont [1]{#1}%
\providecommand \href@noop [0]{\@secondoftwo}%
\providecommand \href [0]{\begingroup \@sanitize@url \@href}%
\providecommand \@href[1]{\@@startlink{#1}\@@href}%
\providecommand \@@href[1]{\endgroup#1\@@endlink}%
\providecommand \@sanitize@url [0]{\catcode `\\12\catcode `\$12\catcode
  `\&12\catcode `\#12\catcode `\^12\catcode `\_12\catcode `\%12\relax}%
\providecommand \@@startlink[1]{}%
\providecommand \@@endlink[0]{}%
\providecommand \url  [0]{\begingroup\@sanitize@url \@url }%
\providecommand \@url [1]{\endgroup\@href {#1}{\urlprefix }}%
\providecommand \urlprefix  [0]{URL }%
\providecommand \Eprint [0]{\href }%
\providecommand \doibase [0]{http://dx.doi.org/}%
\providecommand \selectlanguage [0]{\@gobble}%
\providecommand \bibinfo  [0]{\@secondoftwo}%
\providecommand \bibfield  [0]{\@secondoftwo}%
\providecommand \translation [1]{[#1]}%
\providecommand \BibitemOpen [0]{}%
\providecommand \bibitemStop [0]{}%
\providecommand \bibitemNoStop [0]{.\EOS\space}%
\providecommand \EOS [0]{\spacefactor3000\relax}%
\providecommand \BibitemShut  [1]{\csname bibitem#1\endcsname}%
\let\auto@bib@innerbib\@empty
\bibitem [{\citenamefont {Liu}\ \emph {et~al.}(2000)\citenamefont {Liu},
  \citenamefont {Zhang}, \citenamefont {Mao}, \citenamefont {Zhu},
  \citenamefont {Yang}, \citenamefont {Chan},\ and\ \citenamefont
  {Sheng}}]{Liu_SCIENCE_2000}%
  \BibitemOpen
  \bibfield  {author} {\bibinfo {author} {\bibfnamefont {Z.}~\bibnamefont
  {Liu}}, \bibinfo {author} {\bibfnamefont {X.}~\bibnamefont {Zhang}}, \bibinfo
  {author} {\bibfnamefont {Y.}~\bibnamefont {Mao}}, \bibinfo {author}
  {\bibfnamefont {Y.~Y.}\ \bibnamefont {Zhu}}, \bibinfo {author} {\bibfnamefont
  {Z.}~\bibnamefont {Yang}}, \bibinfo {author} {\bibfnamefont {C.~T.}\
  \bibnamefont {Chan}}, \ and\ \bibinfo {author} {\bibfnamefont
  {P.}~\bibnamefont {Sheng}},\ }\href {\doibase 10.1126/science.289.5485.1734}
  {\bibfield  {journal} {\bibinfo  {journal} {Science}\ }\textbf {\bibinfo
  {volume} {289}},\ \bibinfo {pages} {1734} (\bibinfo {year}
  {2000})}\BibitemShut {NoStop}%
\bibitem [{\citenamefont {Zhu}\ \emph {et~al.}(2011)\citenamefont {Zhu},
  \citenamefont {Christensen}, \citenamefont {Jung}, \citenamefont
  {Martin-Moreno}, \citenamefont {Yin}, \citenamefont {Fok}, \citenamefont
  {Zhang},\ and\ \citenamefont {Garcia-Vidal}}]{Zhu_NAT-PHYS_2011}%
  \BibitemOpen
  \bibfield  {author} {\bibinfo {author} {\bibfnamefont {J.}~\bibnamefont
  {Zhu}}, \bibinfo {author} {\bibfnamefont {J.}~\bibnamefont {Christensen}},
  \bibinfo {author} {\bibfnamefont {J.}~\bibnamefont {Jung}}, \bibinfo {author}
  {\bibfnamefont {L.}~\bibnamefont {Martin-Moreno}}, \bibinfo {author}
  {\bibfnamefont {X.}~\bibnamefont {Yin}}, \bibinfo {author} {\bibfnamefont
  {L.}~\bibnamefont {Fok}}, \bibinfo {author} {\bibfnamefont {X.}~\bibnamefont
  {Zhang}}, \ and\ \bibinfo {author} {\bibfnamefont {F.~J.}\ \bibnamefont
  {Garcia-Vidal}},\ }\href {\doibase 10.1038/nphys1804} {\bibfield  {journal}
  {\bibinfo  {journal} {Nat Phys}\ }\textbf {\bibinfo {volume} {7}},\ \bibinfo
  {pages} {52} (\bibinfo {year} {2011})}\BibitemShut {NoStop}%
\bibitem [{\citenamefont {Zhang}, \citenamefont {Xia},\ and\ \citenamefont
  {Fang}(2011)}]{Zhang_PRL_2011}%
  \BibitemOpen
  \bibfield  {author} {\bibinfo {author} {\bibfnamefont {S.}~\bibnamefont
  {Zhang}}, \bibinfo {author} {\bibfnamefont {C.}~\bibnamefont {Xia}}, \ and\
  \bibinfo {author} {\bibfnamefont {N.}~\bibnamefont {Fang}},\ }\href {\doibase
  10.1103/PhysRevLett.106.024301} {\bibfield  {journal} {\bibinfo  {journal}
  {Phys Rev Lett}\ }\textbf {\bibinfo {volume} {106}},\ \bibinfo {pages}
  {024301} (\bibinfo {year} {2011})}\BibitemShut {NoStop}%
\bibitem [{\citenamefont {Lemoult}\ \emph {et~al.}(2013)\citenamefont
  {Lemoult}, \citenamefont {Kaina}, \citenamefont {Fink},\ and\ \citenamefont
  {Lerosey}}]{Lemoult_NAT-PHYS_2013}%
  \BibitemOpen
  \bibfield  {author} {\bibinfo {author} {\bibfnamefont {F.}~\bibnamefont
  {Lemoult}}, \bibinfo {author} {\bibfnamefont {N.}~\bibnamefont {Kaina}},
  \bibinfo {author} {\bibfnamefont {M.}~\bibnamefont {Fink}}, \ and\ \bibinfo
  {author} {\bibfnamefont {G.}~\bibnamefont {Lerosey}},\ }\href {\doibase
  10.1038/nphys2480} {\bibfield  {journal} {\bibinfo  {journal} {Nat Phys}\
  }\textbf {\bibinfo {volume} {9}},\ \bibinfo {pages} {55} (\bibinfo {year}
  {2013})}\BibitemShut {NoStop}%
\bibitem [{\citenamefont {Hladky-Hennion}\ \emph {et~al.}(2013)\citenamefont
  {Hladky-Hennion}, \citenamefont {Vasseur}, \citenamefont {Haw}, \citenamefont
  {Croënne}, \citenamefont {Haumesser},\ and\ \citenamefont
  {Norris}}]{Hladky_APL_2013}%
  \BibitemOpen
  \bibfield  {author} {\bibinfo {author} {\bibfnamefont {A.-C.}\ \bibnamefont
  {Hladky-Hennion}}, \bibinfo {author} {\bibfnamefont {J.~O.}\ \bibnamefont
  {Vasseur}}, \bibinfo {author} {\bibfnamefont {G.}~\bibnamefont {Haw}},
  \bibinfo {author} {\bibfnamefont {C.}~\bibnamefont {Croënne}}, \bibinfo
  {author} {\bibfnamefont {L.}~\bibnamefont {Haumesser}}, \ and\ \bibinfo
  {author} {\bibfnamefont {A.~N.}\ \bibnamefont {Norris}},\ }\href {\doibase
  10.1063/1.4801642} {\bibfield  {journal} {\bibinfo  {journal} {Appl Phys
  Lett}\ }\textbf {\bibinfo {volume} {102}},\ \bibinfo {pages} {144103}
  (\bibinfo {year} {2013})}\BibitemShut {NoStop}%
\bibitem [{\citenamefont {Zhu}\ and\ \citenamefont
  {Semperlotti}(2016)}]{Zhu_PRL_2016}%
  \BibitemOpen
  \bibfield  {author} {\bibinfo {author} {\bibfnamefont {H.}~\bibnamefont
  {Zhu}}\ and\ \bibinfo {author} {\bibfnamefont {F.}~\bibnamefont
  {Semperlotti}},\ }\href {\doibase 10.1103/PhysRevLett.117.034302} {\bibfield
  {journal} {\bibinfo  {journal} {Phys Rev Lett}\ }\textbf {\bibinfo {volume}
  {117}},\ \bibinfo {pages} {034302} (\bibinfo {year} {2016})}\BibitemShut
  {NoStop}%
\bibitem [{\citenamefont {Addouche}\ \emph {et~al.}(2014)\citenamefont
  {Addouche}, \citenamefont {Al-Lethawe}, \citenamefont {Elayouch},\ and\
  \citenamefont {Khelif}}]{Addouche_APL-ADV_2014}%
  \BibitemOpen
  \bibfield  {author} {\bibinfo {author} {\bibfnamefont {M.}~\bibnamefont
  {Addouche}}, \bibinfo {author} {\bibfnamefont {M.~A.}\ \bibnamefont
  {Al-Lethawe}}, \bibinfo {author} {\bibfnamefont {A.}~\bibnamefont
  {Elayouch}}, \ and\ \bibinfo {author} {\bibfnamefont {A.}~\bibnamefont
  {Khelif}},\ }\href {\doibase 10.1063/1.4901909} {\bibfield  {journal}
  {\bibinfo  {journal} {AIP Adv}\ }\textbf {\bibinfo {volume} {4}},\ \bibinfo
  {pages} {124303} (\bibinfo {year} {2014})}\BibitemShut {NoStop}%
\bibitem [{\citenamefont {Gao}, \citenamefont {Gao},\ and\ \citenamefont
  {Zhang}(2016)}]{Gao_APL_2016}%
  \BibitemOpen
  \bibfield  {author} {\bibinfo {author} {\bibfnamefont {Z.}~\bibnamefont
  {Gao}}, \bibinfo {author} {\bibfnamefont {F.}~\bibnamefont {Gao}}, \ and\
  \bibinfo {author} {\bibfnamefont {B.}~\bibnamefont {Zhang}},\ }\href
  {\doibase 10.1063/1.4940906} {\bibfield  {journal} {\bibinfo  {journal} {Appl
  Phys Lett}\ }\textbf {\bibinfo {volume} {108}},\ \bibinfo {pages} {041105}
  (\bibinfo {year} {2016})}\BibitemShut {NoStop}%
\bibitem [{\citenamefont {Kaina}\ \emph {et~al.}(2016)\citenamefont {Kaina},
  \citenamefont {Causier}, \citenamefont {Bourlier}, \citenamefont {Fink},
  \citenamefont {Berthelot},\ and\ \citenamefont {Lerosey}}]{Kaina_arXiv_2016}%
  \BibitemOpen
  \bibfield  {author} {\bibinfo {author} {\bibfnamefont {N.}~\bibnamefont
  {Kaina}}, \bibinfo {author} {\bibfnamefont {A.}~\bibnamefont {Causier}},
  \bibinfo {author} {\bibfnamefont {Y.}~\bibnamefont {Bourlier}}, \bibinfo
  {author} {\bibfnamefont {M.}~\bibnamefont {Fink}}, \bibinfo {author}
  {\bibfnamefont {T.}~\bibnamefont {Berthelot}}, \ and\ \bibinfo {author}
  {\bibfnamefont {G.}~\bibnamefont {Lerosey}},\ }\href
  {https://arxiv.org/abs/1604.08117} {\bibfield  {journal} {\bibinfo  {journal}
  {arXiv:1604.08117v1}\ } (\bibinfo {year} {2016})}\BibitemShut {NoStop}%
\bibitem [{\citenamefont {Pal}\ and\ \citenamefont
  {Ruzzene}(2017)}]{Pal_NJoP_2017}%
  \BibitemOpen
  \bibfield  {author} {\bibinfo {author} {\bibfnamefont {R.~K.}\ \bibnamefont
  {Pal}}\ and\ \bibinfo {author} {\bibfnamefont {M.}~\bibnamefont {Ruzzene}},\
  }\href {\doibase 10.1088/1367-2630/aa56a2} {\bibfield  {journal} {\bibinfo
  {journal} {New J Phys}\ }\textbf {\bibinfo {volume} {19}},\ \bibinfo {pages}
  {025001} (\bibinfo {year} {2017})}\BibitemShut {NoStop}%
\bibitem [{\citenamefont {Chaunsali}, \citenamefont {Chen},\ and\ \citenamefont
  {Yang}(2018)}]{Chaunsali_arXiv_2018}%
  \BibitemOpen
  \bibfield  {author} {\bibinfo {author} {\bibfnamefont {R.}~\bibnamefont
  {Chaunsali}}, \bibinfo {author} {\bibfnamefont {C.-W.}\ \bibnamefont {Chen}},
  \ and\ \bibinfo {author} {\bibfnamefont {J.}~\bibnamefont {Yang}},\ }\href
  {https://arxiv.org/abs/1806.00655} {\bibfield  {journal} {\bibinfo  {journal}
  {arXiv:1806.00655 [cond-mat.mes-hall]}\ } (\bibinfo {year}
  {2018})}\BibitemShut {NoStop}%
\bibitem [{\citenamefont {Tsakmakidis}, \citenamefont {Boardman},\ and\
  \citenamefont {Hess}(2007)}]{Tsakmakidis_NATURE_2007}%
  \BibitemOpen
  \bibfield  {author} {\bibinfo {author} {\bibfnamefont {K.~L.}\ \bibnamefont
  {Tsakmakidis}}, \bibinfo {author} {\bibfnamefont {A.~D.}\ \bibnamefont
  {Boardman}}, \ and\ \bibinfo {author} {\bibfnamefont {O.}~\bibnamefont
  {Hess}},\ }\href {\doibase 10.1038/nature06285} {\bibfield  {journal}
  {\bibinfo  {journal} {Nature}\ }\textbf {\bibinfo {volume} {450}},\ \bibinfo
  {pages} {397} (\bibinfo {year} {2007})}\BibitemShut {NoStop}%
\bibitem [{\citenamefont {Zhu}\ \emph {et~al.}(2013)\citenamefont {Zhu},
  \citenamefont {Chen}, \citenamefont {Zhu}, \citenamefont {Garcia-Vidal},
  \citenamefont {Yin}, \citenamefont {Zhang},\ and\ \citenamefont
  {Zhang}}]{Zhu_SCIREP_2013}%
  \BibitemOpen
  \bibfield  {author} {\bibinfo {author} {\bibfnamefont {J.}~\bibnamefont
  {Zhu}}, \bibinfo {author} {\bibfnamefont {Y.}~\bibnamefont {Chen}}, \bibinfo
  {author} {\bibfnamefont {X.}~\bibnamefont {Zhu}}, \bibinfo {author}
  {\bibfnamefont {F.~J.}\ \bibnamefont {Garcia-Vidal}}, \bibinfo {author}
  {\bibfnamefont {X.}~\bibnamefont {Yin}}, \bibinfo {author} {\bibfnamefont
  {W.}~\bibnamefont {Zhang}}, \ and\ \bibinfo {author} {\bibfnamefont
  {X.}~\bibnamefont {Zhang}},\ }\href {\doibase 10.1038/srep01728} {\bibfield
  {journal} {\bibinfo  {journal} {Sci Rep}\ }\textbf {\bibinfo {volume} {3}},\
  \bibinfo {pages} {1728} (\bibinfo {year} {2013})}\BibitemShut {NoStop}%
\bibitem [{\citenamefont {Kr\"odel}, \citenamefont {Thom\'e},\ and\
  \citenamefont {Daraio}(2015)}]{Kroedel_EML_2015}%
  \BibitemOpen
  \bibfield  {author} {\bibinfo {author} {\bibfnamefont {S.}~\bibnamefont
  {Kr\"odel}}, \bibinfo {author} {\bibfnamefont {N.}~\bibnamefont {Thom\'e}}, \
  and\ \bibinfo {author} {\bibfnamefont {C.}~\bibnamefont {Daraio}},\ }\href
  {\doibase 10.1016/j.eml.2015.05.004} {\bibfield  {journal} {\bibinfo
  {journal} {Extreme Mech Lett}\ }\textbf {\bibinfo {volume} {4}},\ \bibinfo
  {pages} {111} (\bibinfo {year} {2015})}\BibitemShut {NoStop}%
\bibitem [{\citenamefont {Zhao}, \citenamefont {Li},\ and\ \citenamefont
  {Zhu}(2015)}]{Zhao_SCIREP_2015}%
  \BibitemOpen
  \bibfield  {author} {\bibinfo {author} {\bibfnamefont {D.-G.}\ \bibnamefont
  {Zhao}}, \bibinfo {author} {\bibfnamefont {Y.}~\bibnamefont {Li}}, \ and\
  \bibinfo {author} {\bibfnamefont {X.-F.}\ \bibnamefont {Zhu}},\ }\href
  {\doibase 10.1038/srep09376} {\bibfield  {journal} {\bibinfo  {journal} {Sci
  Rep}\ }\textbf {\bibinfo {volume} {5}},\ \bibinfo {pages} {9376} (\bibinfo
  {year} {2015})}\BibitemShut {NoStop}%
\bibitem [{\citenamefont {Cardella}, \citenamefont {Celli},\ and\ \citenamefont
  {Gonella}(2016)}]{Cardella_SMS_2016}%
  \BibitemOpen
  \bibfield  {author} {\bibinfo {author} {\bibfnamefont {D.}~\bibnamefont
  {Cardella}}, \bibinfo {author} {\bibfnamefont {P.}~\bibnamefont {Celli}}, \
  and\ \bibinfo {author} {\bibfnamefont {S.}~\bibnamefont {Gonella}},\ }\href
  {\doibase 10.1088/0964-1726/25/8/085017} {\bibfield  {journal} {\bibinfo
  {journal} {Smart Mater Struct}\ }\textbf {\bibinfo {volume} {25}},\ \bibinfo
  {pages} {085017} (\bibinfo {year} {2016})}\BibitemShut {NoStop}%
\bibitem [{\citenamefont {Tian}\ and\ \citenamefont
  {Yu}(2017)}]{Tian_SCIREP_2017}%
  \BibitemOpen
  \bibfield  {author} {\bibinfo {author} {\bibfnamefont {Z.}~\bibnamefont
  {Tian}}\ and\ \bibinfo {author} {\bibfnamefont {L.}~\bibnamefont {Yu}},\
  }\href {\doibase 10.1038/srep40004} {\bibfield  {journal} {\bibinfo
  {journal} {Sci Rep}\ }\textbf {\bibinfo {volume} {7}},\ \bibinfo {pages}
  {40004} (\bibinfo {year} {2017})}\BibitemShut {NoStop}%
\bibitem [{\citenamefont {Colombi}\ \emph {et~al.}(2017)\citenamefont
  {Colombi}, \citenamefont {Ageeva}, \citenamefont {Smith}, \citenamefont
  {Clare}, \citenamefont {Patel}, \citenamefont {Clark}, \citenamefont
  {Colquitt}, \citenamefont {Roux}, \citenamefont {Guenneau},\ and\
  \citenamefont {Craster}}]{Colombi_SCIREP_2017}%
  \BibitemOpen
  \bibfield  {author} {\bibinfo {author} {\bibfnamefont {A.}~\bibnamefont
  {Colombi}}, \bibinfo {author} {\bibfnamefont {V.}~\bibnamefont {Ageeva}},
  \bibinfo {author} {\bibfnamefont {R.~J.}\ \bibnamefont {Smith}}, \bibinfo
  {author} {\bibfnamefont {A.}~\bibnamefont {Clare}}, \bibinfo {author}
  {\bibfnamefont {R.}~\bibnamefont {Patel}}, \bibinfo {author} {\bibfnamefont
  {M.}~\bibnamefont {Clark}}, \bibinfo {author} {\bibfnamefont
  {D.}~\bibnamefont {Colquitt}}, \bibinfo {author} {\bibfnamefont
  {P.}~\bibnamefont {Roux}}, \bibinfo {author} {\bibfnamefont {S.}~\bibnamefont
  {Guenneau}}, \ and\ \bibinfo {author} {\bibfnamefont {R.~V.}\ \bibnamefont
  {Craster}},\ }\href {\doibase 10.1038/s41598-017-07151-6} {\bibfield
  {journal} {\bibinfo  {journal} {Sci. Rep.}\ }\textbf {\bibinfo {volume}
  {7}},\ \bibinfo {pages} {6750} (\bibinfo {year} {2017})}\BibitemShut
  {NoStop}%
\bibitem [{\citenamefont {Guo}, \citenamefont {Hettich},\ and\ \citenamefont
  {Dekorsy}(2017)}]{Guo_NJoP_2017}%
  \BibitemOpen
  \bibfield  {author} {\bibinfo {author} {\bibfnamefont {Y.}~\bibnamefont
  {Guo}}, \bibinfo {author} {\bibfnamefont {M.}~\bibnamefont {Hettich}}, \ and\
  \bibinfo {author} {\bibfnamefont {T.}~\bibnamefont {Dekorsy}},\ }\href
  {\doibase 10.1088/1367-2630/aa5703} {\bibfield  {journal} {\bibinfo
  {journal} {New J Phys}\ }\textbf {\bibinfo {volume} {19}},\ \bibinfo {pages}
  {013029} (\bibinfo {year} {2017})}\BibitemShut {NoStop}%
\bibitem [{\citenamefont {Wu}\ \emph {et~al.}(2008)\citenamefont {Wu},
  \citenamefont {Huang}, \citenamefont {Tsai},\ and\ \citenamefont
  {Wu}}]{Wu_2008}%
  \BibitemOpen
  \bibfield  {author} {\bibinfo {author} {\bibfnamefont {T.-T.}\ \bibnamefont
  {Wu}}, \bibinfo {author} {\bibfnamefont {Z.-G.}\ \bibnamefont {Huang}},
  \bibinfo {author} {\bibfnamefont {T.-C.}\ \bibnamefont {Tsai}}, \ and\
  \bibinfo {author} {\bibfnamefont {T.-C.}\ \bibnamefont {Wu}},\ }\href
  {\doibase 10.1063/1.2970992} {\bibfield  {journal} {\bibinfo  {journal} {Appl
  Phys Lett}\ }\textbf {\bibinfo {volume} {93}},\ \bibinfo {pages} {111902}
  (\bibinfo {year} {2008})}\BibitemShut {NoStop}%
\bibitem [{\citenamefont {Pennec}\ \emph {et~al.}(2008)\citenamefont {Pennec},
  \citenamefont {Djafari-Rouhani}, \citenamefont {Larabi}, \citenamefont
  {Vasseur},\ and\ \citenamefont {Hladky-Hennion}}]{Pennec_2008}%
  \BibitemOpen
  \bibfield  {author} {\bibinfo {author} {\bibfnamefont {Y.}~\bibnamefont
  {Pennec}}, \bibinfo {author} {\bibfnamefont {B.}~\bibnamefont
  {Djafari-Rouhani}}, \bibinfo {author} {\bibfnamefont {H.}~\bibnamefont
  {Larabi}}, \bibinfo {author} {\bibfnamefont {J.~O.}\ \bibnamefont {Vasseur}},
  \ and\ \bibinfo {author} {\bibfnamefont {A.~C.}\ \bibnamefont
  {Hladky-Hennion}},\ }\href {\doibase 10.1103/PhysRevB.78.104105} {\bibfield
  {journal} {\bibinfo  {journal} {Phys Rev B}\ }\textbf {\bibinfo {volume}
  {78}},\ \bibinfo {pages} {104105} (\bibinfo {year} {2008})}\BibitemShut
  {NoStop}%
\bibitem [{\citenamefont {Oudich}\ \emph {et~al.}(2011)\citenamefont {Oudich},
  \citenamefont {Senesi}, \citenamefont {Assouar}, \citenamefont {Ruzenne},
  \citenamefont {Sun}, \citenamefont {Vincent}, \citenamefont {Hou},\ and\
  \citenamefont {Wu}}]{Oudich_PRB_2011}%
  \BibitemOpen
  \bibfield  {author} {\bibinfo {author} {\bibfnamefont {M.}~\bibnamefont
  {Oudich}}, \bibinfo {author} {\bibfnamefont {M.}~\bibnamefont {Senesi}},
  \bibinfo {author} {\bibfnamefont {M.~B.}\ \bibnamefont {Assouar}}, \bibinfo
  {author} {\bibfnamefont {M.}~\bibnamefont {Ruzenne}}, \bibinfo {author}
  {\bibfnamefont {J.-H.}\ \bibnamefont {Sun}}, \bibinfo {author} {\bibfnamefont
  {B.}~\bibnamefont {Vincent}}, \bibinfo {author} {\bibfnamefont
  {Z.}~\bibnamefont {Hou}}, \ and\ \bibinfo {author} {\bibfnamefont {T.-T.}\
  \bibnamefont {Wu}},\ }\href {\doibase 10.1103/PhysRevB.84.165136} {\bibfield
  {journal} {\bibinfo  {journal} {Phys Rev B}\ }\textbf {\bibinfo {volume}
  {84}},\ \bibinfo {pages} {165136} (\bibinfo {year} {2011})}\BibitemShut
  {NoStop}%
\bibitem [{\citenamefont {Assouar}\ \emph {et~al.}(2012)\citenamefont
  {Assouar}, \citenamefont {Senesi}, \citenamefont {Oudich}, \citenamefont
  {Ruzzene},\ and\ \citenamefont {Hou}}]{Assouar_APL_2012}%
  \BibitemOpen
  \bibfield  {author} {\bibinfo {author} {\bibfnamefont {M.~B.}\ \bibnamefont
  {Assouar}}, \bibinfo {author} {\bibfnamefont {M.}~\bibnamefont {Senesi}},
  \bibinfo {author} {\bibfnamefont {M.}~\bibnamefont {Oudich}}, \bibinfo
  {author} {\bibfnamefont {M.}~\bibnamefont {Ruzzene}}, \ and\ \bibinfo
  {author} {\bibfnamefont {Z.}~\bibnamefont {Hou}},\ }\href {\doibase
  10.1063/1.4764072} {\bibfield  {journal} {\bibinfo  {journal} {Appl Phys
  Lett}\ }\textbf {\bibinfo {volume} {101}},\ \bibinfo {pages} {173505}
  (\bibinfo {year} {2012})}\BibitemShut {NoStop}%
\bibitem [{\citenamefont {Casadei}\ \emph {et~al.}(2012)\citenamefont
  {Casadei}, \citenamefont {Delpero}, \citenamefont {Bergamini}, \citenamefont
  {Ermanni},\ and\ \citenamefont {Ruzzene}}]{Casadei_JAP_2012}%
  \BibitemOpen
  \bibfield  {author} {\bibinfo {author} {\bibfnamefont {F.}~\bibnamefont
  {Casadei}}, \bibinfo {author} {\bibfnamefont {T.}~\bibnamefont {Delpero}},
  \bibinfo {author} {\bibfnamefont {A.}~\bibnamefont {Bergamini}}, \bibinfo
  {author} {\bibfnamefont {P.}~\bibnamefont {Ermanni}}, \ and\ \bibinfo
  {author} {\bibfnamefont {M.}~\bibnamefont {Ruzzene}},\ }\href {\doibase
  10.1063/1.4752468} {\bibfield  {journal} {\bibinfo  {journal} {J Appl Phys}\
  }\textbf {\bibinfo {volume} {112}},\ \bibinfo {eid} {064902} (\bibinfo {year}
  {2012})}\BibitemShut {NoStop}%
\bibitem [{\citenamefont {Achaoui}\ \emph {et~al.}(2013)\citenamefont
  {Achaoui}, \citenamefont {Laude}, \citenamefont {Benchabane},\ and\
  \citenamefont {Khelif}}]{Achaoui_JAP_2013}%
  \BibitemOpen
  \bibfield  {author} {\bibinfo {author} {\bibfnamefont {Y.}~\bibnamefont
  {Achaoui}}, \bibinfo {author} {\bibfnamefont {V.}~\bibnamefont {Laude}},
  \bibinfo {author} {\bibfnamefont {S.}~\bibnamefont {Benchabane}}, \ and\
  \bibinfo {author} {\bibfnamefont {A.}~\bibnamefont {Khelif}},\ }\href
  {\doibase 10.1063/1.4820928} {\bibfield  {journal} {\bibinfo  {journal} {J
  Appl Phys}\ }\textbf {\bibinfo {volume} {114}},\ \bibinfo {pages} {104503}
  (\bibinfo {year} {2013})}\BibitemShut {NoStop}%
\bibitem [{\citenamefont {Rupin}\ \emph {et~al.}(2014)\citenamefont {Rupin},
  \citenamefont {Lemoult}, \citenamefont {Lerosey},\ and\ \citenamefont
  {Roux}}]{Rupin_PRL_2014}%
  \BibitemOpen
  \bibfield  {author} {\bibinfo {author} {\bibfnamefont {M.}~\bibnamefont
  {Rupin}}, \bibinfo {author} {\bibfnamefont {F.}~\bibnamefont {Lemoult}},
  \bibinfo {author} {\bibfnamefont {G.}~\bibnamefont {Lerosey}}, \ and\
  \bibinfo {author} {\bibfnamefont {P.}~\bibnamefont {Roux}},\ }\href {\doibase
  10.1103/PhysRevLett.112.234301} {\bibfield  {journal} {\bibinfo  {journal}
  {Phys Rev Lett}\ }\textbf {\bibinfo {volume} {112}},\ \bibinfo {pages}
  {234301} (\bibinfo {year} {2014})}\BibitemShut {NoStop}%
\bibitem [{\citenamefont {Pourabolghasem}\ \emph {et~al.}(2014)\citenamefont
  {Pourabolghasem}, \citenamefont {Khelif}, \citenamefont {Mohammadi},
  \citenamefont {Eftekhar},\ and\ \citenamefont
  {Adibi}}]{Pourabolghasem_JAP_2014}%
  \BibitemOpen
  \bibfield  {author} {\bibinfo {author} {\bibfnamefont {R.}~\bibnamefont
  {Pourabolghasem}}, \bibinfo {author} {\bibfnamefont {A.}~\bibnamefont
  {Khelif}}, \bibinfo {author} {\bibfnamefont {S.}~\bibnamefont {Mohammadi}},
  \bibinfo {author} {\bibfnamefont {A.~A.}\ \bibnamefont {Eftekhar}}, \ and\
  \bibinfo {author} {\bibfnamefont {A.}~\bibnamefont {Adibi}},\ }\href
  {\doibase 10.1063/1.4887115} {\bibfield  {journal} {\bibinfo  {journal} {J
  Appl Phys}\ }\textbf {\bibinfo {volume} {116}},\ \bibinfo {pages} {013514}
  (\bibinfo {year} {2014})}\BibitemShut {NoStop}%
\bibitem [{\citenamefont {Celli}\ and\ \citenamefont
  {Gonella}(2015)}]{Celli_APL_2015}%
  \BibitemOpen
  \bibfield  {author} {\bibinfo {author} {\bibfnamefont {P.}~\bibnamefont
  {Celli}}\ and\ \bibinfo {author} {\bibfnamefont {S.}~\bibnamefont
  {Gonella}},\ }\href {\doibase 10.1063/1.4929566} {\bibfield  {journal}
  {\bibinfo  {journal} {Appl Phys Lett}\ }\textbf {\bibinfo {volume} {107}},\
  \bibinfo {pages} {081901} (\bibinfo {year} {2015})}\BibitemShut {NoStop}%
\bibitem [{\citenamefont {Wu}, \citenamefont {Harne},\ and\ \citenamefont
  {Wang}(2016)}]{Wu_JIMSS_2015}%
  \BibitemOpen
  \bibfield  {author} {\bibinfo {author} {\bibfnamefont {Z.}~\bibnamefont
  {Wu}}, \bibinfo {author} {\bibfnamefont {R.~L.}\ \bibnamefont {Harne}}, \
  and\ \bibinfo {author} {\bibfnamefont {K.~W.}\ \bibnamefont {Wang}},\ }\href
  {\doibase 10.1177/1045389X15586451} {\bibfield  {journal} {\bibinfo
  {journal} {J Intel Mat Syst Struct}\ }\textbf {\bibinfo {volume} {27}},\
  \bibinfo {pages} {1189} (\bibinfo {year} {2016})}\BibitemShut {NoStop}%
\bibitem [{\citenamefont {Lee}\ \emph {et~al.}(2016)\citenamefont {Lee},
  \citenamefont {Kang}, \citenamefont {Keum}, \citenamefont {Ahmed},
  \citenamefont {Rogers}, \citenamefont {Ferreira}, \citenamefont {Kim},\ and\
  \citenamefont {Min}}]{Lee_SCIREP_2016}%
  \BibitemOpen
  \bibfield  {author} {\bibinfo {author} {\bibfnamefont {S.}~\bibnamefont
  {Lee}}, \bibinfo {author} {\bibfnamefont {B.}~\bibnamefont {Kang}}, \bibinfo
  {author} {\bibfnamefont {H.}~\bibnamefont {Keum}}, \bibinfo {author}
  {\bibfnamefont {N.}~\bibnamefont {Ahmed}}, \bibinfo {author} {\bibfnamefont
  {J.~A.}\ \bibnamefont {Rogers}}, \bibinfo {author} {\bibfnamefont {P.~M.}\
  \bibnamefont {Ferreira}}, \bibinfo {author} {\bibfnamefont {S.}~\bibnamefont
  {Kim}}, \ and\ \bibinfo {author} {\bibfnamefont {B.}~\bibnamefont {Min}},\
  }\href {\doibase 10.1038/srep27621} {\bibfield  {journal} {\bibinfo
  {journal} {Sci Rep}\ }\textbf {\bibinfo {volume} {6}},\ \bibinfo {pages}
  {27621} (\bibinfo {year} {2016})}\BibitemShut {NoStop}%
\bibitem [{\citenamefont {Memoli}\ \emph {et~al.}(2017)\citenamefont {Memoli},
  \citenamefont {Caleap}, \citenamefont {Asakawa}, \citenamefont {Sahoo},
  \citenamefont {Drinkwater},\ and\ \citenamefont
  {Subramanian}}]{Memoli_NATCOMM_2017}%
  \BibitemOpen
  \bibfield  {author} {\bibinfo {author} {\bibfnamefont {G.}~\bibnamefont
  {Memoli}}, \bibinfo {author} {\bibfnamefont {M.}~\bibnamefont {Caleap}},
  \bibinfo {author} {\bibfnamefont {M.}~\bibnamefont {Asakawa}}, \bibinfo
  {author} {\bibfnamefont {D.~R.}\ \bibnamefont {Sahoo}}, \bibinfo {author}
  {\bibfnamefont {B.~W.}\ \bibnamefont {Drinkwater}}, \ and\ \bibinfo {author}
  {\bibfnamefont {S.}~\bibnamefont {Subramanian}},\ }\href {\doibase
  10.1038/ncomms14608} {\bibfield  {journal} {\bibinfo  {journal} {Nat Commun}\
  }\textbf {\bibinfo {volume} {8}},\ \bibinfo {pages} {14608} (\bibinfo {year}
  {2017})}\BibitemShut {NoStop}%
\bibitem [{\citenamefont {Goffaux}\ and\ \citenamefont
  {Vigneron}(2001)}]{Goffaux_PRB_2001}%
  \BibitemOpen
  \bibfield  {author} {\bibinfo {author} {\bibfnamefont {C.}~\bibnamefont
  {Goffaux}}\ and\ \bibinfo {author} {\bibfnamefont {J.~P.}\ \bibnamefont
  {Vigneron}},\ }\href {\doibase 10.1103/PhysRevB.64.075118} {\bibfield
  {journal} {\bibinfo  {journal} {Phys Rev B}\ }\textbf {\bibinfo {volume}
  {64}},\ \bibinfo {pages} {075118} (\bibinfo {year} {2001})}\BibitemShut
  {NoStop}%
\bibitem [{\citenamefont {Romero-Garc\`ia}\ \emph {et~al.}(2013)\citenamefont
  {Romero-Garc\`ia}, \citenamefont {Lagarrigue}, \citenamefont {Groby},
  \citenamefont {Richoux},\ and\ \citenamefont
  {Tournat}}]{Romero-Garcia_JPD_2013}%
  \BibitemOpen
  \bibfield  {author} {\bibinfo {author} {\bibfnamefont {V.}~\bibnamefont
  {Romero-Garc\`ia}}, \bibinfo {author} {\bibfnamefont {C.}~\bibnamefont
  {Lagarrigue}}, \bibinfo {author} {\bibfnamefont {J.-P.}\ \bibnamefont
  {Groby}}, \bibinfo {author} {\bibfnamefont {O.}~\bibnamefont {Richoux}}, \
  and\ \bibinfo {author} {\bibfnamefont {V.}~\bibnamefont {Tournat}},\ }\href
  {\doibase 10.1088/0022-3727/46/30/305108} {\bibfield  {journal} {\bibinfo
  {journal} {J Phys D: Appl Phys}\ }\textbf {\bibinfo {volume} {46}},\ \bibinfo
  {pages} {305108} (\bibinfo {year} {2013})}\BibitemShut {NoStop}%
\bibitem [{\citenamefont {Thota}, \citenamefont {Li},\ and\ \citenamefont
  {Wang}(2017)}]{Thota_PRB_2017}%
  \BibitemOpen
  \bibfield  {author} {\bibinfo {author} {\bibfnamefont {M.}~\bibnamefont
  {Thota}}, \bibinfo {author} {\bibfnamefont {S.}~\bibnamefont {Li}}, \ and\
  \bibinfo {author} {\bibfnamefont {K.~W.}\ \bibnamefont {Wang}},\ }\href
  {\doibase 10.1103/PhysRevB.95.064307} {\bibfield  {journal} {\bibinfo
  {journal} {Phys Rev B}\ }\textbf {\bibinfo {volume} {95}},\ \bibinfo {pages}
  {064307} (\bibinfo {year} {2017})}\BibitemShut {NoStop}%
\bibitem [{\citenamefont {Williams}\ \emph {et~al.}(2015)\citenamefont
  {Williams}, \citenamefont {Roux}, \citenamefont {Rupin},\ and\ \citenamefont
  {Kuperman}}]{Williams_PRB_2015}%
  \BibitemOpen
  \bibfield  {author} {\bibinfo {author} {\bibfnamefont {E.~G.}\ \bibnamefont
  {Williams}}, \bibinfo {author} {\bibfnamefont {P.}~\bibnamefont {Roux}},
  \bibinfo {author} {\bibfnamefont {M.}~\bibnamefont {Rupin}}, \ and\ \bibinfo
  {author} {\bibfnamefont {W.~A.}\ \bibnamefont {Kuperman}},\ }\href {\doibase
  10.1103/PhysRevB.91.104307} {\bibfield  {journal} {\bibinfo  {journal} {Phys
  Rev B}\ }\textbf {\bibinfo {volume} {91}},\ \bibinfo {pages} {104307}
  (\bibinfo {year} {2015})}\BibitemShut {NoStop}%
\bibitem [{\citenamefont {Bilal}\ and\ \citenamefont
  {Hussein}(2013)}]{Bilal_APL_2013}%
  \BibitemOpen
  \bibfield  {author} {\bibinfo {author} {\bibfnamefont {O.~R.}\ \bibnamefont
  {Bilal}}\ and\ \bibinfo {author} {\bibfnamefont {M.~I.}\ \bibnamefont
  {Hussein}},\ }\href {\doibase 10.1063/1.4820796} {\bibfield  {journal}
  {\bibinfo  {journal} {Appl Phys Lett}\ }\textbf {\bibinfo {volume} {103}},\
  \bibinfo {pages} {111901} (\bibinfo {year} {2013})}\BibitemShut {NoStop}%
\bibitem [{\citenamefont {Banerjee}, \citenamefont {Das},\ and\ \citenamefont
  {Calius}(2017)}]{Banerjee_JAP_2017}%
  \BibitemOpen
  \bibfield  {author} {\bibinfo {author} {\bibfnamefont {A.}~\bibnamefont
  {Banerjee}}, \bibinfo {author} {\bibfnamefont {R.}~\bibnamefont {Das}}, \
  and\ \bibinfo {author} {\bibfnamefont {E.~P.}\ \bibnamefont {Calius}},\
  }\href {\doibase 10.1063/1.4998446} {\bibfield  {journal} {\bibinfo
  {journal} {J Appl Phys}\ }\textbf {\bibinfo {volume} {122}},\ \bibinfo
  {pages} {075101} (\bibinfo {year} {2017})}\BibitemShut {NoStop}%
\bibitem [{\citenamefont {Coffy}\ \emph {et~al.}(2015)\citenamefont {Coffy},
  \citenamefont {Lavergne}, \citenamefont {Addouche}, \citenamefont
  {Euphrasie}, \citenamefont {Vairac},\ and\ \citenamefont
  {Khelif}}]{Coffy_JAP_2015}%
  \BibitemOpen
  \bibfield  {author} {\bibinfo {author} {\bibfnamefont {E.}~\bibnamefont
  {Coffy}}, \bibinfo {author} {\bibfnamefont {T.}~\bibnamefont {Lavergne}},
  \bibinfo {author} {\bibfnamefont {M.}~\bibnamefont {Addouche}}, \bibinfo
  {author} {\bibfnamefont {S.}~\bibnamefont {Euphrasie}}, \bibinfo {author}
  {\bibfnamefont {P.}~\bibnamefont {Vairac}}, \ and\ \bibinfo {author}
  {\bibfnamefont {A.}~\bibnamefont {Khelif}},\ }\href {\doibase
  10.1063/1.4936836} {\bibfield  {journal} {\bibinfo  {journal} {J Appl Phys}\
  }\textbf {\bibinfo {volume} {118}},\ \bibinfo {pages} {214902} (\bibinfo
  {year} {2015})}\BibitemShut {NoStop}%
\bibitem [{\citenamefont {Oh}, \citenamefont {Seung},\ and\ \citenamefont
  {Kim}(2016)}]{Oh_APL_2016}%
  \BibitemOpen
  \bibfield  {author} {\bibinfo {author} {\bibfnamefont {J.~H.}\ \bibnamefont
  {Oh}}, \bibinfo {author} {\bibfnamefont {H.~M.}\ \bibnamefont {Seung}}, \
  and\ \bibinfo {author} {\bibfnamefont {Y.~Y.}\ \bibnamefont {Kim}},\ }\href
  {\doibase 10.1063/1.4943095} {\bibfield  {journal} {\bibinfo  {journal} {Appl
  Phys Lett}\ }\textbf {\bibinfo {volume} {108}},\ \bibinfo {pages} {093501}
  (\bibinfo {year} {2016})}\BibitemShut {NoStop}%
\bibitem [{\citenamefont {Thorp}, \citenamefont {Ruzzene},\ and\ \citenamefont
  {Baz}(2001)}]{Thorp_SMS_2001}%
  \BibitemOpen
  \bibfield  {author} {\bibinfo {author} {\bibfnamefont {O.}~\bibnamefont
  {Thorp}}, \bibinfo {author} {\bibfnamefont {M.}~\bibnamefont {Ruzzene}}, \
  and\ \bibinfo {author} {\bibfnamefont {A.}~\bibnamefont {Baz}},\ }\href
  {\doibase 10.1088/0964-1726/10/5/314} {\bibfield  {journal} {\bibinfo
  {journal} {Smart Mater Struct}\ }\textbf {\bibinfo {volume} {10}},\ \bibinfo
  {pages} {979} (\bibinfo {year} {2001})}\BibitemShut {NoStop}%
\bibitem [{\citenamefont {Sainidou}, \citenamefont {Stefanou},\ and\
  \citenamefont {Modinos}(2005)}]{Sainidou_PRL_2005}%
  \BibitemOpen
  \bibfield  {author} {\bibinfo {author} {\bibfnamefont {R.}~\bibnamefont
  {Sainidou}}, \bibinfo {author} {\bibfnamefont {N.}~\bibnamefont {Stefanou}},
  \ and\ \bibinfo {author} {\bibfnamefont {A.}~\bibnamefont {Modinos}},\ }\href
  {\doibase 10.1103/PhysRevLett.94.205503} {\bibfield  {journal} {\bibinfo
  {journal} {Phys Rev Lett}\ }\textbf {\bibinfo {volume} {94}},\ \bibinfo
  {pages} {205503} (\bibinfo {year} {2005})}\BibitemShut {NoStop}%
\bibitem [{\citenamefont {Richoux}, \citenamefont {Morand},\ and\ \citenamefont
  {Simon}(2009)}]{Richoux_2009}%
  \BibitemOpen
  \bibfield  {author} {\bibinfo {author} {\bibfnamefont {O.}~\bibnamefont
  {Richoux}}, \bibinfo {author} {\bibfnamefont {E.}~\bibnamefont {Morand}}, \
  and\ \bibinfo {author} {\bibfnamefont {L.}~\bibnamefont {Simon}},\
  }\href@noop {} {\bibfield  {journal} {\bibinfo  {journal} {Ann Phys}\
  }\textbf {\bibinfo {volume} {324}},\ \bibinfo {pages} {1983} (\bibinfo {year}
  {2009})}\BibitemShut {NoStop}%
\bibitem [{\citenamefont {Lin}(1996)}]{Lin_AMR_1996}%
  \BibitemOpen
  \bibfield  {author} {\bibinfo {author} {\bibfnamefont {Y.-K.}\ \bibnamefont
  {Lin}},\ }\href@noop {} {\bibfield  {journal} {\bibinfo  {journal} {Appl Mech
  Rev}\ }\textbf {\bibinfo {volume} {49}},\ \bibinfo {pages} {57} (\bibinfo
  {year} {1996})}\BibitemShut {NoStop}%
\bibitem [{bla()}]{blah}%
  \BibitemOpen
  \href@noop {} {}\bibinfo {note} {An alternative to increasing $\delta h$ is
  increasing the value of the conical mass that slides along each beam. We
  found this to have a qualitatively similar effect to increasing $\delta
  h$.}\BibitemShut {Stop}%
\end{thebibliography}

%


\clearpage
\widetext

\setcounter{figure}{0}
\setcounter{page}{1}
\setcounter{section}{0}
\renewcommand{\thefigure}{S\arabic{figure}}
\renewcommand{\thetable}{S\arabic{table}}
\renewcommand{\theequation}{S\arabic{equation}}
\renewcommand{\thesection}{S\arabic{section}}
\renewcommand{\thesubsection}{S\arabic{section}.\arabic{subsection}}
\renewcommand{\thepage}{S\arabic{page}}
\makeatletter

\section*{Supplemental Material for ``Bandgap widening by disorder in rainbow metamaterials''}

\section{Additional information on the setup and experimental results}

\subsection{Details on the experimental setup}

The specimen is a strip cut out from a gray acrylonitrile butadiene styrene (ABS) baseplate (LEGO\textsuperscript{\textregistered}, Item 10701) and features 12$\times$5 telescopic resonators arranged in a square lattice architecture. The strip is $5.7\,\mathrm{cm}$ wide and the distance between the clamped ends is approximately $46.3\,\mathrm{cm}$. Each resonator is characterized by a rod/pillar (LEGO\textsuperscript{\textregistered}, Elem. ID 395726) and a conical brick (LEGO\textsuperscript{\textregistered}, Elem. ID 4518029) in prismatic contact. The conical brick can be slid up and down the rod to tune the pillar's natural frequency; the rod-brick contact is strong enough for the brick to hold its position throughout the tests.  Note that the pillars are attached to the baseplate through frictional contact as well (anchoring the base of the rod to one of the protuberances, \emph{studs}, of the baseplate). This allows for an agile reconfiguration of the brick arrangement. The standing wave excitation signals are transmitted to the structure through an electromechanical shaker (Br\"uel \& Kj\ae r Type $4810$) and a stinger. The out of plane velocity time histories of points on the back-side of the plate belonging to a pre-determined grid are recorded via a 3D Scanning Laser Doppler Vibrometer (3D-SLDV, Polytec PSV-400-3D). The acquisition is performed in the frequency domain (the Fast Fourier Transform is performed automatically within the PSV acquisition system). To eliminate non-repeatable noisy features from the response, measurements are repeated 10 times and averaged at each measurement location. As far as the vibrometer channel is concerned, we select a $5\,\mathrm{V}$ range and DC coupling. We acquire in the $0\mbox{--}5\,\mathrm{kHz}$ frequency range, and we concentrate on the $0\mbox{--}600\,\mathrm{Hz}$ band when postprocessing the data. The sampling frequency is $f_s=12.8\,\mathrm{kHz}$ and the number of FFT lines is 3200, resulting in a frequency resolution of $1.5625\,\mathrm{Hz}$. The selected velocity decoder is the digital VD-08-$10\,\mathrm{mm/s/V}$, that allows acquisitions up to $20\,\mathrm{kHz}$. The excitation is a pseudorandom waveform with maximum amplitude of $500\,\mathrm{mV}$. The excitation signal is amplified using a Br\"uel \& Kj\ae r Type $2718$ Power Amplifier, with gain set to $30\,\mathrm{dB}$.

\begin{figure} [!htb]
\centering
\includegraphics[scale=1.4]{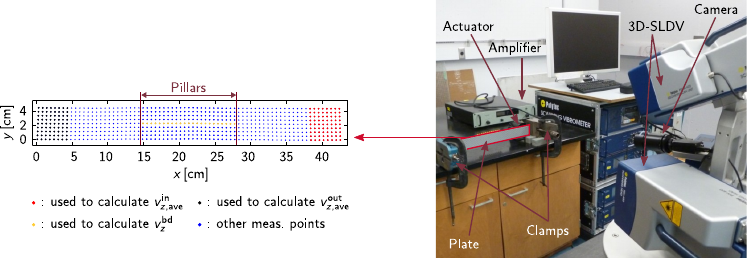}
\caption{Measurement grid (left, where the groups of points are used to calculate the quantities reported in the legend). Picture of the experimental setup (right).}
\label{fig:setupS}
\end{figure}

The grid of measurement points is shown in Fig.~\ref{fig:setupS}.
The legend shows which sets of points are used to calculate certain quantities of interest. Note that the transmissibility is measured as $v_{z,\mathrm{ave}}^{\mathrm{out}}/v_{z,\mathrm{ave}}^{\mathrm{in}}$, while $v_{z,\mathrm{ave}}^{\mathrm{bd}}$ is used to reconstruct the dispersion relation of each configuration. In particular, the reconstruction operation is performed by taking the frequency-space data for all points highlighted in yellow, and performing a 1D discrete Fourier transform. This yields frequency-wavenumber spectral maps. The dispersion branches are extracted by tracing the maxima of the spectral function at each frequency.

\begin{figure} [!htb]
\centering
\includegraphics[scale=1.38]{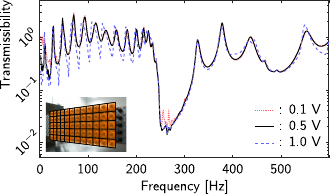}
\caption{Response of a uniform M1 architecture to three different loading amplitudes.}
\label{fig:nonlin}
\end{figure}

Our specimens are extremely flexible and are made of a polymeric material. Moreover, since the pillars are attached to the baseplate via frictional contacts, it is reasonable to assume that, above a certain amplitude of excitation, these contacts can lead to nonlinearities in the response. To rule out nonlinearities from our explanations, we compare the responses of the same architecture featuring only M1 resonators to excitations at different amplitudes. The transmissibility plots for three loading amplitudes ($0.1\,\mathrm{V}$, $0.5\,\mathrm{V}$---the amplitude used for all experiments throughout this article---and $1.0\,\mathrm{V}$) are superimposed in Fig.~\ref{fig:nonlin}.
We can see that the only difference between the $0.1\,\mathrm{V}$ case and the $0.5\,\mathrm{V}$ case is represented by the morphology of the bandgap---with the $0.1\,\mathrm{V}$ case being characterized by a jagged ``bottom''. With respect to the other two cases, the high amplitude ($1.0\,\mathrm{V}$) one is characterized by a different response both before and after the bandgap. We believe this amplitude-dependent behavior to be indeed due to a mix of material and geometric nonlinearities that are triggered when the load is larger than a certain threshold. However, we exclude that nonlinearities affect our observations on bandgap widening, due to the fact that the amplitude of excitation does not significantly affect the extent of the bandgap, whose sharp onset and sloping end are unchanged.

\clearpage
\subsection{Response of uniform architectures}

\begin{figure} [!htb]
\centering
\includegraphics[scale=1.4]{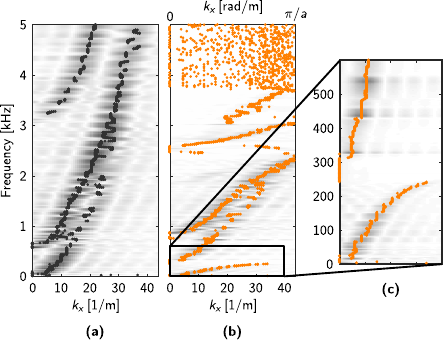}
\caption{(a), (b) Experimentally-reconstructed dispersion relation for a plate with no pillars, and for a plate with 12$\times$5 identical M1-type pillars, respectively. (c) Low-frequency detail of (b), highlighting the hybridization bandgap of interest. Mode shapes for the case with M1 pillars, measured along the centerline of the plate strip and recorded at frequencies before, within and after the hybridization bandgap.}
\label{fig:bandsM1}
\end{figure}

In this section, we report additional results on the response of uniform architectures. The comparison between the experimentally-reconstructed dispersion curves of a plate with no pillars and of a plate with 12$\times$5 identical M1-type pillars is shown in Fig.~\ref{fig:bandsM1}(a,b).
Fig.~\ref{fig:bandsM1}(c) shows the frequency range of interest, where only one mode exists. Fig.~\ref{fig:bandsM1}(a,b), on the other hand, show a much wider frequency range, where multiple modes are present. While we don't have an explanation for all the modes in this range, we can see that the influence of the bricks is also significant at higher frequencies.

\begin{figure} [!htb]
\centering
\includegraphics[scale=1.4]{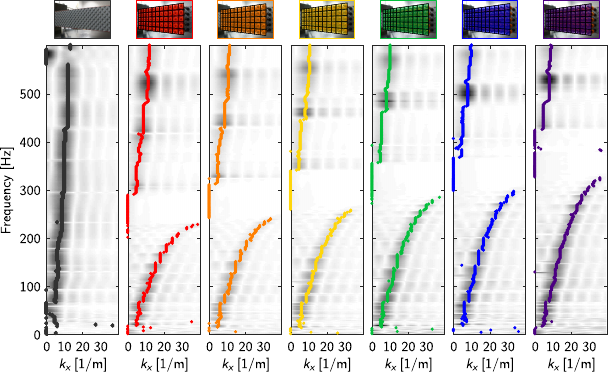}
\caption{From left to right: reconstructed band diagrams of configurations featuring no pillars, pillars of the T, M1, M2, M3, M4, M5 type. The dots on the band diagrams are the maxima of the spectral function at each frequency.}
\label{fig:bandsALL}
\end{figure}

In Fig.~\ref{fig:bandsALL}, we report the low-frequency reconstructed band diagrams of all uniform configurations. 
We can see that the bandgap consistently shifts towards higher frequencies as we lower the conical brick along the pillar. Here, we define the bandgap as that frequency range where $k_x=0$. While identifying the bandgaps is trivial for the T, M1, M2, M3, M4 configurations, things are not so clear for M5. This configuration seems to feature a wide gap split by a horizontal mode, that could be due to the mechanics of the pillar when the conical brick is located near its base. Note that this phenomenon is not captured numerically, when the brick is approximated as a point mass. For the reader's convenience, all bandgap ranges extracted from these experimental results are tabulated in Table~\ref{tab:resexp}. 
\begin{table}[!htb]
\begin{center}
\begin{ruledtabular}
\begin{tabular}{c|c|c|c|c|c|c}
	Configuration & T & M1 & M2 & M3 & M4 & M5 \\ \hline
	$f_\mathrm{onset}$ (Hz) & 228 & 242 & 259 & 284 & 298 & 327 \\ \hline
	$f_\mathrm{end}$ (Hz) & 289 & 311 & 339 & 355 & 366 & 423*
\end{tabular}
\end{ruledtabular}
\caption{\label{tab:resexp}
Experimental bandgap ranges. *: This bandgap is split by a mode of unknown origin at around 380 Hz.}
\end{center}
\end{table}

\clearpage
\subsection{More responses of heterogeneous architectures}

\begin{figure} [!htb]
\centering
\includegraphics[scale=1.4]{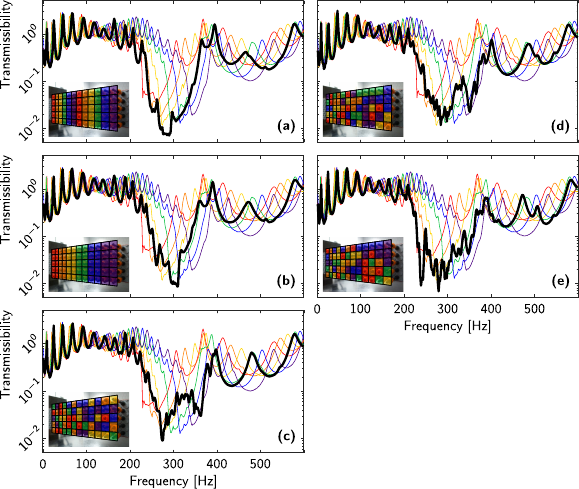}
\caption{Influence of the spatial arrangement of heterogeneously tuned resonators on wave attenuation. In all cases, 10/60 resonators are programmed to the brick configurations T, M1, M2, M3, M4 and M5. (a), (b) Transmissibilities for graded and (c), (d), (e) spatially randomized configurations, marked by thick black lines; thin color-coded lines refer to uniformly tuned monochromatic configurations (see color coding in Fig.~1).}
\label{fig:heterog6type}
\end{figure}

\begin{figure} [!htb]
\centering
\includegraphics[scale=1.4]{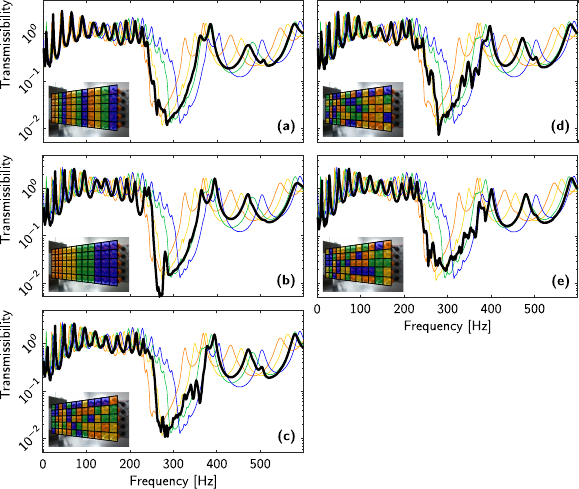}
\caption{Influence of the spatial arrangement of heterogeneously tuned resonators on wave attenuation. In all cases, 15/60 resonators are programmed to the brick configurations M1, M2, M3, and M4. (a), (b) Transmissibilities for graded and (c), (d), (e) spatially randomized configurations, marked by thick black lines.}
\label{fig:heterog4type}
\end{figure}

In this section, we report additional results related to the attenuation capabilities of architectures featuring graded and disordered spatial arrangements of heterogeneous resonators. In the article, we reported on various architectures displaying 10 resonators of each of the following types: T, M1, M2, M3, M4, M5. All the results we obtained with these sets of resonators are shown in Fig.~\ref{fig:heterog6type}. 
In Fig.~\ref{fig:heterog4type}, we show that similar considerations can be made for architectures comprising 15 M1, 15 M2, 15 M3 and 15 M4 resonators.
We can see that bandgaps for the graded architectures, shown in Figs.~\ref{fig:heterog4type}a-b, span two of the reference bandgaps (yellow and green) and present a similar morphology (sharp onset and sloping end). On the other hand, the bandgaps of configurations featuring disordered brick arrangements are wider than their graded counterparts, spanning three or four individual bandgaps as shown in Figs.~\ref{fig:heterog4type}c-e, and also present different morphological characteristics (they have a ``jagged'' profile, while also being less deep).

\begin{figure} [!htb]
\centering
\includegraphics[scale=1.4]{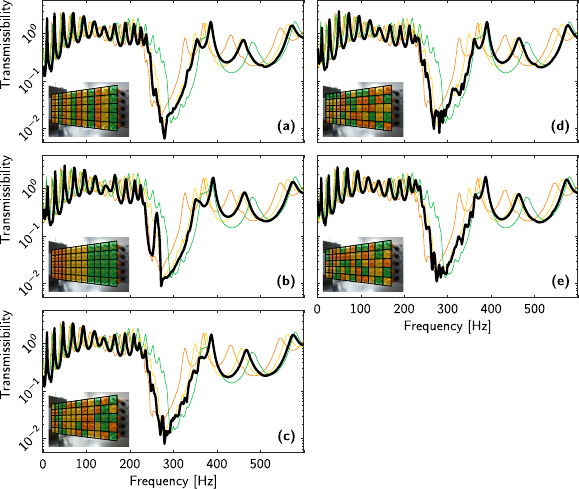}
\caption{Influence of the spatial arrangement of heterogeneously tuned resonators on wave attenuation. In all cases, 20/60 resonators are programmed to the brick configurations M1, M2 and M3. (a), (b) Transmissibilities for graded and (c), (d), (e) spatially randomized configurations, marked by thick black lines.}
\label{fig:heterog3type}
\end{figure}

In Fig.~\ref{fig:heterog3type}, we show the response of architectures comprising 20 M1, 20 M2 and 20 M3 resonators. Similar considerations as in the previous case apply, and we still observe a slight widening due to randomization.

\begin{figure} [!htb]
\centering
\includegraphics[scale=1.4]{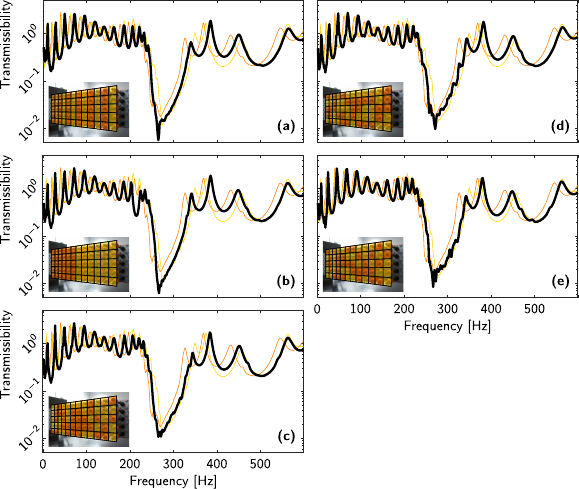}
\caption{Influence of the spatial arrangement of heterogeneously tuned resonators on wave attenuation. In all cases, 30/60 resonators are programmed to the brick configurations M1 and M2. (a), (b) Transmissibilities for graded and (c), (d), (e) spatially randomized configurations, marked by thick black lines.}
\label{fig:heterog2type}
\end{figure}

In Fig.~\ref{fig:heterog2type}, we show the response of architectures comprising 30 M1 and 30 M2 resonators.
As we decrease the degree of heterogeneity among resonator characteristics, we can see that the results for graded and disordered architectures do not differ much. Even though it is challenging to comment on the bandgap width, we are able to see that the bandgaps in Figs.~\ref{fig:heterog2type}a,b have a more regular morphology than those in Figs.~\ref{fig:heterog2type}c-e.

\clearpage
\section{Additional information on the model and numerical results}

\subsection{Material properties for the numerical model}

Table~\ref{tab:prop} summarizes the material properties used for numerical simulation of the metamaterial system. The Young's modulus and density are denoted by $E$ and $\rho$, the conical mass of each pillar with $m_0$, Poisson's ratio with $\nu$ and the value of structural damping with $\eta$. The value of Poisson's ratio was chosen based on known data for ABS, the material from which the plate and pillars are made. The value of structural damping was chosen such that the simulated and measured transfer functions are of the same order of magnitude. Other parameters in Table~\ref{tab:prop} are based on measurements. 
\begin{table}[!htb]
\begin{center}
\caption{\label{tab:prop}
Material properties used in numerical simulation of the metamaterial system.}
\begin{ruledtabular}
\begin{tabular}{c|c|c|c|c|c|c}
	$E_\textnormal{plate}$ & $\rho_\textnormal{plate}$ & $E_\textnormal{pillar}$ & $\rho_\textnormal{pillar}$ & $m_0$ & $\nu$ & $\eta$ \\ \hline
	5.5~GPa & 8.16$\times10^{-4}$~g/mm$^3$ & 27.2~GPa & 11.0$\times10^{-4}$~g/mm$^3$ & 0.219~g & 0.35 & 0.015~Ns$^2$/m
\end{tabular}
\end{ruledtabular}
\end{center}
\end{table}

\clearpage
\subsection{Computation of dispersion relations}
\label{sec:FEdispersion}

To compute the diversion curves of our metamaterial system, we define as our unit cell a portion of the plate that contains one column of 5 pillars. The units cell has a length of 11.25~mm along the $x$ axis and a width of 56.25~mm along the $y$ axis. The top and bottom edges of the unit cell are free, and Bloch boundary conditions are applied to the right and left edges. This unit cell is appropriate for capturing the dispersion properties of the uniform configurations due to the locally-resonant nature of the bandgap. 

\begin{figure} [!htb]
\centering
\includegraphics[scale=1.4]{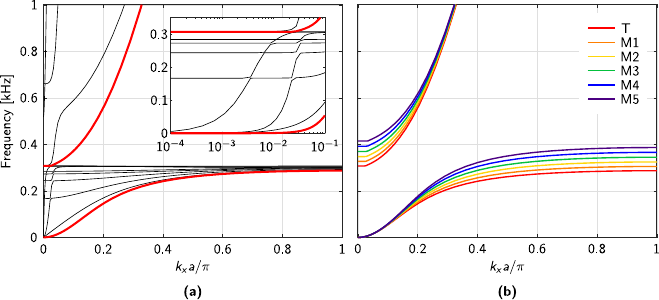}
\caption{Computed dispersion diagrams of the metamaterial system. (a) The first 14 modes of the tall configuration. The branches highlighted in red thick curves correspond to modes that have a non-negligible out-of-plane motion along the centreline. The inset magnifies the long-wavelength portion of the dispersion diagram. (b) Dispersion diagrams of the 6 uniform configurations (cf. Fig.~\ref{fig:bandsALL}). The length of the unit cell is $a=11.25$~mm.}
\label{fig:FEdispersion}
\end{figure}

Fig.~\ref{fig:FEdispersion}a shows the dispersion diagram of the tall configuration (T) containing all the modes up to 1~kHz (the first 14 modes of the unit cell). An inspection of the modes up to 600~Hz (not reported for brevity) reveals that the majority of mode shapes exhibit considerable twist/torsional motion where the centreline (along the $x$ axis) remains relatively motionless. While the modes with torsional motion do exist, they are not excited in our metamaterial system because of the two fixed boundary conditions as well as the mid-plane excitation. Out experimental reconstruction of the dispersion relation in Figs.~2b and \ref{fig:bandsALL} confirms this claim. Keeping only the mode shapes with non-negligible out-of-plane motion along the centreline, we are left with two branches in the dispersion diagram (modes 1 and 11). These branches are highlighted in thick red curves in Fig.~\ref{fig:FEdispersion}a, and agree with measurements of Fig.~\ref{fig:bandsALL}. Fig.~\ref{fig:FEdispersion}b shows the computed dispersion diagrams for the 6 uniform configurations. 

\clearpage
\subsection{Detailed comparison of bandgaps}

\begin{figure} [!htb]
\centering
\includegraphics[scale=1.4]{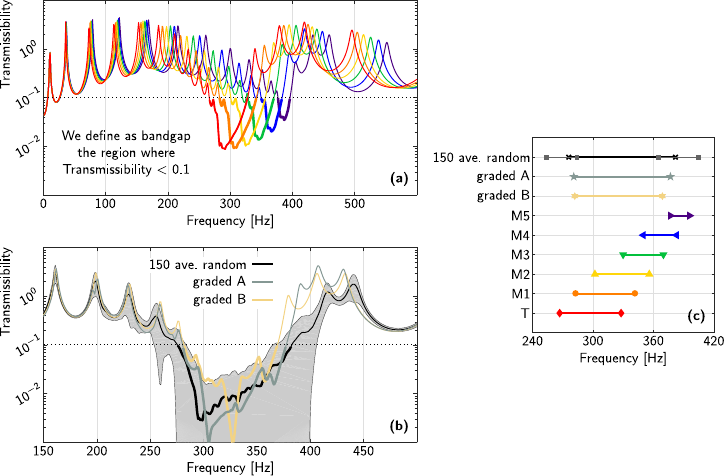}
\caption{Comparison of numerical transmissibility curves and bandgap widths for different spatial arrangements of pillars. (a) Transmissibility curves for the six uniform configurations. (b) Transmissibility curves for random and graded configurations. (c) Comparison of the bandgap widths for uniform, graded and random configurations.}
\label{fig:gapdetails}
\end{figure}

We compare the bandgaps of different configurations: 6 uniform, 2 graded and random. 
Fig.~\ref{fig:gapdetails}a shows the transmissibility curves for the 6 homogeneous pillar populations (T, M1, M2, M3, M4, M5) introduced in Fig.~1. 
We have defined the bandgap as the frequency range where less than 1\% percent of wave energy is transmitted through the plate. This corresponds to the portion of trnamissibility curves that lies below the value of 0.1 (see Fig.~\ref{fig:gapdetails}a). 
Comparison of Fig.~\ref{fig:gapdetails}a to their measured counterparts in Fig.~2d shows that the numerical model closely reproduces the qualitative features of the transmissibility curves. The same is observed when comparing the bandsgaps of the graded arrangements in Fig.~\ref{fig:gapdetails}b (simulated) and Figs.3a,3b (measured). 

Fig.~\ref{fig:gapdetails}c shows how the bandgap evolves according to the spatial arrangement of the pillars. As the conical mass is slid down the pillars in uniform configurations (from T to M5), the bandgap shifts to higher frequencies and becomes narrower. The shift to higher frequencies occurs because the effective inertia of the sliding mass becomes smaller. The narrowing effect occurs because the relative amplitude of oscillations of the pillar decreases, resulting in weaker coupling between the pillars. The bandgaps of the graded and random arrangements have similar widths, but all three are much wider than bandgaps of the uniform arrangements. It is important to note that the bandgap of the random arrangement is the widest in the \emph{ensemble-average} sense, meaning that individual realizations could have narrower bandgaps.

\begin{figure} [!htb]
\centering
\includegraphics[scale=1.4]{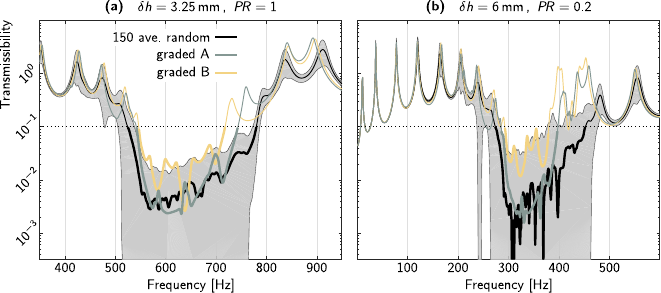}
\caption{Comparison of the transmissibilities for those parameters where the bandgap of the random configurations is the widest. The gray area corresponds to the standard deviation of the transmissibility within the ensemble. Panel (a) corresponds to Fig.~4c and panel (b) to Fig.~4d. In both cases, the random configuration has the widest bandgap on average.}
\label{fig:widest}
\end{figure}

Our parametric study of bandgaps revealed that stiffening the plate (Fig.~4c) and increasing the spacing between the resonance frequencies of pillars (Fig.~4d) widens the bandgaps for graded and random configurations. Fig.~\ref{fig:widest} compares the transmissibilities of graded and random configuration for those parameter values that resulted in the widest bandgaps. As reported in Fig.~4, the random configuration has a wider bandgap than either of the graded configurations. When we consider the standard deviation of the transmissibilities within the ensemble of random configurations, we note that individual realizations may have bandgaps with a similar width to those of graded configurations (see Fig.~\ref{fig:widest}b). 

\begin{figure} [!htb]
\centering
\includegraphics[scale=1.4]{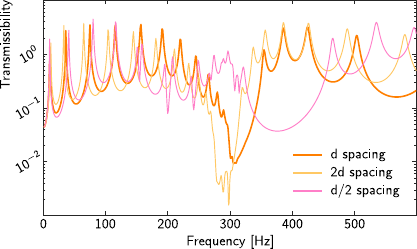}
\caption{Influence of the spacing between pillars (along $x$) on the transmissibility of the M1 configuration. The experimental setup has a spacing of $d=11.25$~mm.}
\label{fig:spacing}
\end{figure}

Fig.~\ref{fig:spacing} shows the influence of the spacing between pillars (along the $x$ axis) on the transmissibility of the M1 configuration -- similar results are obtained for other uniform configurations. Increasing the spacing $d$ makes the coupling between pillars weaker, which results in a narrower bandgap. A similar effect is obtained by softening the plate, as reported in Fig.~4 for random and graded configurations. 

\clearpage
\subsection{Spatial profile of the response}

We have focused mainly on the influence of the spatial arrangement of pillars on the transmissiblity of the metamaterial. In this section, we consider their influence on the spatial profile of the response, as quantified by the velocity amplitudes at the center line of the plate along the $x$ axis. Fig.2c shows the measured spatial profile of the plate for the uniform arrangement M1 -- similar spatial profiles are obtained for other uniform arrangements. Here, we compare the spatial profiles for the graded and random arrangements. 

\begin{figure} [!htb]
\centering
\includegraphics[scale=1.4]{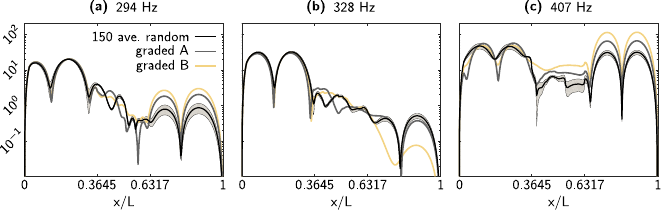}
\caption{Comparison of the spatial profiles of the response for graded and random arrangements. The gray area shows the standard deviation of the black curve. The start and end of the region containing the pillars correspond to $x/L=0.36$ and $x/L=0.63$, respectively, where $L=463$~mm is the total length of the plate. }
\label{fig:spatialS}
\end{figure}

Fig.~\ref{fig:spatialS} shows the spatial profiles at three frequencies near the bandgaps -- see Fig.~\ref{fig:gapdetails}b for the corresponding transmissibility curves. As expected, the spatial profile of the response is highly dependent on the frequency, specifically within and after the region populated by the pillars. 
Within the shared bandgap frequencies (Figs.~\ref{fig:spatialS}a and~b), the three configurations have a somewhat similar spatial profile in the region containing the pillars, and their relative response amplitudes on the receiver side agrees well with their transmissibilities in Fig.~\ref{fig:gapdetails}b. 
At 407~Hz (Fig.~\ref{fig:spatialS}c), the graded configurations are already well outside their bandgap. Accordingly, its velocity response is markedly lower than the response of the graded configurations within and after the locally resonant region. 

\end{document}